\documentclass[aip,apl,reprint,a4paper,amsmath]{revtex4-1}
\usepackage{graphicx}%
\usepackage{dcolumn}%
\usepackage{bm}%

\begin{document}

\title{Variation of the metallic content of Focused Electron Beam Induced Deposition of Cobalt}
\author{L Bernau}
\email{laurent.bernau@empa.ch}
\author{M Gabureac}
\author{I Utke}
\affiliation{Swiss Federal Laboratories for Materials Testing and Research (EMPA), CH-3602 Thun, Switzerland}

\begin{abstract}
Cobalt-containing deposits from Cobalt carbonyl are experimentally produced and their composition is measured. The Cobalt concentration is found to be readily tunable between 20 and 70 at.\% by variation of the dwell time. The variations in metallic concentration are explained by co-deposition of hydrocarbons present in the chamber background pressure. 
\end{abstract}
\maketitle
\section{Introduction}
Focused electron-beam induced deposition (FEBID) is a promising technology for the maskless deposition of metallic structures on a nano-scale \cite{utkereview,vandorpreview,randolphreview}. The electron-beam is scanned in the presence of adsorbed organo-metallic precursors, leading to dissociation of the molecules and desorption of the volatile by-products, thus defining local, metal-containing structures in a range defined by the beam diameter and the primary electron interaction volume from which the emitted electrons, secondaries and backscattered, arises. Deposition of various ferromagnetic metals has been investigated, such as Iron \cite{hochleitner,shimojo,lukasczyk}, Nickel \cite{perentes} or Cobalt \cite{lau,utkespt,utkecross,utkedensity}. Clean magnetic metal deposits were achieved using UHV setups \cite{lukasczyk}, or high beam currents on relatively thick (300 nm) SiO$_2$ insulating films \cite{fernandez}, or beam heating effects \cite{shimojo,utkethermal}. 

However, it is in certain cases desirable to deposit a \emph{composite} nanostructure having the magnetic metal embedded as nanocrystals in a carbonaceous matrix \cite{utkecross,utkethermal} serving for mechanical stability and as oxidation barrier in ambient or liquid atmospheres, for instance in magnetic scanning probe applications \cite{utkespt}. Furthermore, this composite material is of particular interest for the fabrication of sub-micron Hall sensors, as this material exhibits a large extraordinary Hall effect (EHE) \cite{boero}. A precise control of the deposition process is thus required, as it dominantly influences the magnetic sensing properties of the deposit \cite{gabureacMNE}. Here, we present a study of the deposition process in planar deposits obtained from dicobalt-octacarbonyl (Co$_2$(CO)$_8$). While the deposition of cobalt-containing pillars, where the electron-beam is constantly illuminating a single spot on the substrate, has been investigated in depth \cite{lau,utkedensity,utkethermal}, only few studies of the linear or planar deposition of Cobalt, where the beam is raster scanned over a certain area, have been published so far \cite{fernandez}. We find metallic concentrations varying between 20 and 70 at.\% and explain these findings by the co-deposition of hydrocarbons present in the chamber background.

\section{Experimental methods}

The FEBID experiments were carried out at room temperature in a Hitachi S-3600N scanning electron microscope (SEM), equipped with a tungsten filament electron source. Silicon substrates covered by a 200nm-thick SiO$_2$ layer were used as a deposition substrate. Homemade modifications to the SEM allowed placing an internal reservoir for the precursor, connected to a syringe nozzle pointed towards the substrate. Dicobalt octacarbonyl (Co$_2$Co$_8$, CAS 10210-68-1) was used as a precursor molecule. In order to prevent spontaneous decomposition of cobalt carbonyl, the precursor is usually stabilized with 5-10 wt.\% hexane (Sigma-Aldrich GmbH, Buchs, Switzerland). However, the hexane stabilization proves to be a rich source for carbon co-deposition during FEBID. When stored in an atmosphere of argon (VWR International, Dietikon, Switzerland), cobalt carbonyl yields deposits of higher cobalt content when dissociated by FEBID. 

The precursor flux during the experiments was about 4.4$\times$10$^{15}$ molecules/s, as estimated from mass loss measurements. According to our gas flow MC simulations \cite{friedligis}, this translates into 1.5$\times$10$^{17}$ cm$^{-2}$s$^{-1}$ impinging on the FEB irradiated spot. The residual chamber pressure was estimated from pressure measurements carried out without precursor, after degasing of the experimental setup. In the presence of an internal precursor reservoir, the monitored chamber pressure eventually saturates. Deposition was carried out after stabilization of the chamber pressure, and the residual pressure was estimated to ~1$\times$10$^{-5}$ mbar. 

\begin{figure}
\begin{center}
\includegraphics[width=.45\textwidth]{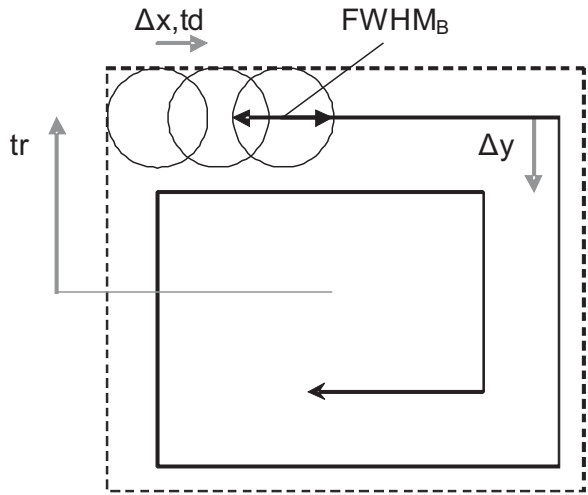}
\end{center}
\caption{Scanning parameters. The rectangular deposits are realized using a serpentine deposition path. $\Delta x$ and $\Delta y$ denote the increment in x and y direction and are identical throughout the experiments. Each pixel is irradiated for a dwell time $t_d$. The refresh time is the time elapsing between to subsequent irradiations of the same pixel and is given by the multiplication of the number of pixels with the dwell time.}
\label{fig:writingstrategy}
\end{figure}

We used the XENOS lithography system to control the electron beam during deposition. The deposition control parameters are the nominal dwell time $t_d$, i.e. the time the beam irradiates a pixel before moving to the next, the inter-pixel distance, defined as the distance between to adjacent pixels irradiated by the beam, and the refresh time $t_r$, which measures the time elapsed before the next iteration of the structure irradiation (see figure \ref{fig:writingstrategy}). Dividing the total irradiation time by the total exposed surface yields the deposition dose in C/cm$^2$. In our experiments, the total deposition dose was kept constant to 10C/cm$^2$, while the dwell time was varied over two orders of magnitude (500ns to 50$\mu$s). The refresh time was always kept $\ge$10ms, to ensure full replenishment with precursor molecules between two irradiation iterations. The inter-pixel distance was set to 30nm. This compares to a beam diameter (FWHM) estimated to 70nm (assessed experimentally using BeamMetr, \cite{babin}), for the acceleration voltage (25kV) and beam current (1nA) used. This leads to an overlap of roughly 2, and the effective dwell time is the double of the nominal dwell time. The total irradiation time per pixel is then the double of the effective dwell time, as the overlap is present in both x- and y-direction.

The deposits were characterized with regard to the composition using energy dispersive x-ray spectroscopy (EDX), using a 3keV probe. The accuracy of the compositional values obtained by this mean was checked to be correct within $\pm$5at.\% by calibration measurements on Cobalt-carbonate (CoCO$_3$). The height of the deposits was assessed by atomic force microscopy (AFM).

\section{Results}

\subsection{Composition}

\begin{figure}
\begin{center}
\includegraphics[width=.45\textwidth]{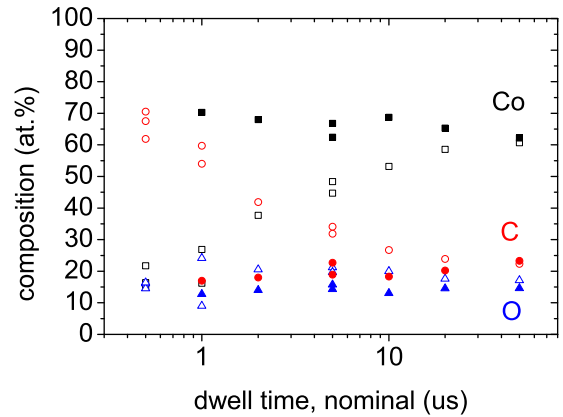}
\end{center}
\caption{Composition of the deposits as measured by EDX for varying dwell times. Filled symbols are for (3$\mu$m)$^2$ structures, open symbols are for (600nm)$^2$ deposits.}
\label{fig:compositionsquar}
\end{figure}

The metallic concentration of the deposits is found to be highly dependent on the dwell time for structures with a width of 600nm, as shown in figure \ref{fig:compositionsquar}. For dwell times below 1$\mu$s, Cobalt concentrations as low as 20at.\% are found, while dwell times of 20$\mu$s or more yield deposits with above 60at.\% Cobalt. For deposits with a characteristic dimension of 3$\mu$m however, the Cobalt concentration is within 60-70at.\% independently of the dwell time used, with a tendency for lower metallic concentrations for higher dwell times. 

The balance of the composition contains Oxygen and Carbon. For the 600nm squares, the Carbon content scales inversely with the Cobalt content, decreasing from about 70at.\% for dwell times of 500ns down to 22at.\% for dwell times of 50$\mu$s, whereas the Oxygen level remains stable at around 20at.\%. For the 3$\mu$m squares, the Oxygen level remains stable within the measurement error at around 15at.\% and a Carbon level of about 20at.\% is found. 

The EDX measurements were performed by focusing a 3keV electron beam on the center of the deposit, so that the escape cone for the X-rays has a radius of well below 50nm. Thus, the discrepancy in the concentrations found for the 600nm squares cannot be attributed to edge effects during the measurement.

\subsection{Deposition rates}

\begin{figure}
\begin{center}
\includegraphics[width=.50\textwidth]{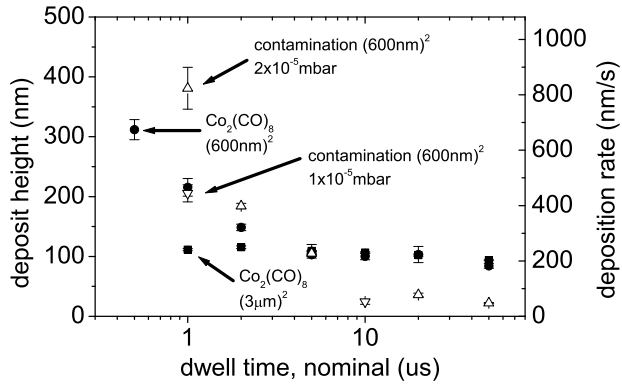}
\end{center}
\caption{Heights of the investigated deposits, for (3$\mu$m)$^2$ (full squares) and (600nm)$^2$ (full circles) Co$_2$(CO)$_8$ deposition. Error bars show spread in height measurements along three adjacent AFM profiles. The contamination deposition was performed at ~2$\times$10$^{-5}$mbar (upwards pointing triangles) and ~1$\times$10$^{-5}$mbar (downwards pointing triangles). The right-hand axis gives the corresponding growth rates.}
\label{fig:heights2}
\end{figure}

The deposit heights are reported in figure \ref{fig:heights2}. The corresponding deposition rates (right-hand axis of figure \ref{fig:heights2}) are obtained by dividing the heights by the total irradiation time per pixel, as defined in the experimental section. As the total irradiation time per pixel, the beam current and the beam diameter are constant throughout all experimental series, the deposit heights and the corresponding deposition rates scale linearly. For the 600$\times$600nm$^2$ squares, the deposition rates are found to increase strongly for low dwell times (below 5$\mu$s), and are nearly constant for dwell times above this threshold. For the 3$\times$3$\mu$m$^2$ squares however, nearly constant deposition rates are found for all dwell times investigated. 

As a comparison, the contamination deposition rates, obtained by depositing 600$\times$600nm$^2$ squares in the absence of any precursor, are shown, for background pressure of 1 and 2 $\times$10$^{-5}$mbar. A strong dependency on the dwell time is found in this case, and the absolute value of the deposition rate at the same dwell time setting is found to depend linearly on the background chamber pressure, i.e. on the amount of hydrocarbon ``precursor'' available for the deposition process.

\section{Discussion}

The growth rate is described by the model \cite{utkereview}:
\begin{eqnarray}
R=\frac{V\sigma f}{t_d}\int\limits_{0}^{t_d}n(t)\,dt
\label{eq:eq1}
\end{eqnarray}
where $V$ is the volume of the decomposed molecule, $\sigma$ is the decomposition cross-section, $f$ is the PE flux and $n(t)$ describes the surface coverage with precursor molecule at a given time. Disregarding surface diffusion, the precursor coverage is described by:
\begin{eqnarray}
\frac{\partial n}{\partial t}=sJ\left(1-\frac{n}{n_0}\right)-\frac{n}{\tau}-\sigma fn
\label{eq:eq2}
\end{eqnarray}
Assuming full replenishment between successive irradiations, the growth rate equation can be solved analytically by integration of equation \ref{eq:eq2}:
\begin{eqnarray}
R=V\sigma f \left[\left(n_{out}-n_{in}\right)\frac{1-exp\left(-k_dt_d\right)}{k_dt_d}+n_{in}\right]
\label{eq:eq3}
\end{eqnarray}
where $n_{out}$ and $n_{in}$ represent the molecular coverage inside and outside the electron beam, and $k_d$ is the time constant of the dissociative depletion process. Similarly, a time constant $k_r$ for the precursor replenishment between irradiations can be introduced.

\begin{subequations}
\label{eq:eq4}
\begin{eqnarray}
k_d=sJ/n_0+1/\tau+\sigma f \\
k_r=sJ/n_0+1/\tau \\
n_{in}=sJ/\left(sJ/n_0+1/\tau+\sigma f\right) \\
n_{out}=sJ/\left(sJ/n_0+1/\tau\right)
\end{eqnarray}
\end{subequations}

\begin{figure}
\begin{center}
\includegraphics[width=.45\textwidth]{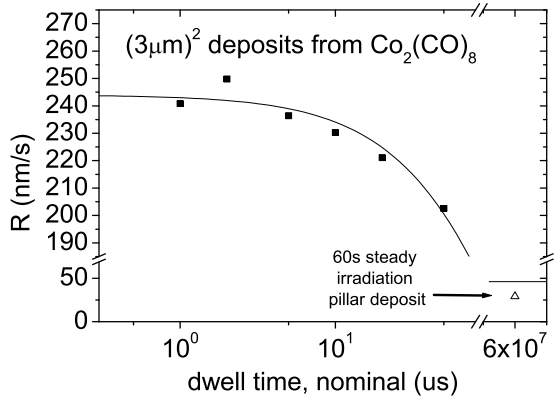}
\end{center}
\caption{Growth rates of (3$\mu$m)$^2$ deposits from Co$_2$(CO)$_8$. The total dose was maintained constant at 10C/cm$^2$. The solid line represents a fit of equation \ref{eq:eq3}. The fit is extended to 60s and the growth rate from a steady-exposure pillar is indicated for comparison.}
\label{fig:R3umCo}
\end{figure}

\begin{figure}
\begin{center}
\includegraphics[width=.45\textwidth]{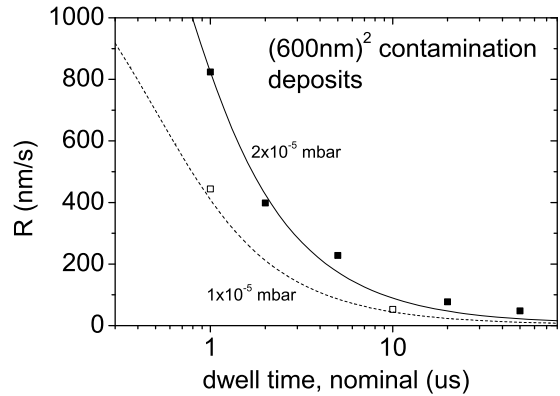}
\end{center}
\caption{Growth rates of (600nm)$^2$ contamination deposits at 1 and 2$\times$10$^{-5}$mbar. The solid line represents a fit of equation \ref{eq:eq3} to the values at 2$\times$10$^{-5}$mbar using polyphenyl ether (PPE) or benzene as example hydrocarbon molecules. Using the fitted parameters, the model growth rates at 1$\times$10$^{-5}$mbar are drawn (dotted line).}
\label{fig:R600nmcont}
\end{figure}

The condition for full replenishment is fulfilled for $k_r\cdot t_r \gg 1$, i.e. when the refresh time is much higher than the time constant for the replenishment process. While the impinging flux of precursor molecules J can be estimated from mass loss measurements and MC simulations, the monolayer coverage $n_0$ approximated using the molecule's size, the sticking coefficient $s$ and the residence time $\tau$ are unknown molecular parameters in the case of Co$_2$(CO)$_8$. Using as a typical value $\tau$=1ms and assuming $s$=1, the refresh time used in our experiments proves sufficient to fulfill full replenishment conditions. The growth rates for the 3$\mu$m$^{2}$ Cobalt carbonyl and the contamination deposits series were fitted using the values given in table \ref{tb:fitconstants} and the model curves are shown in figure \ref{fig:R3umCo} and \ref{fig:R600nmcont} (see table \ref{tb:fitparameters}). Although the growth rate model used (equ. \ref{eq:eq3}) does not account explicitely for surface diffusion of adsorbed molecules, diffusive effect contribute by variations of the effective residence time. 

\begin{table}
\caption{
Model parameters for Co$_2$(CO)$_8$ and two hydrocarbons. Two molecules are taken as example for the fit of the contamination deposits: benzene (C$_{6}$H$_{6}$) is taken as an example of a ``light'' hydrocarbon, as its dissociation under electron exposure has been investigated in \cite{Kunze}; 1,3-Bis(3-phenoxyphenoxy)benzene (C$_{30}$H$_{22}$O$_{4}$) is a polyphenyl ether (PPE) used in the oil diffusion pump and is taken as an example of a ``heavy'' hydrocarbon. 
}
\label{tb:fitconstants}
\begin{ruledtabular}
\begin{tabular}{@{}lcccc}
precursor
&$J$
&$n_0$
&$f$
&$V$
\\
&cm$^{-2}$s$^{-1}$
&nm$^{-2}$
&cm$^{-2}$s$^{-1}$
&nm$^{3}$
\\
Co$_2$(CO)$_8$	
&$1.5\cdot10^{17} \footnotemark[1]$	
&	$2.6\footnotemark[3]$
&$1.7\cdot10^{20}$
&$3.8\cdot10^{-2}\footnotemark[5]$
\\
PPE
&$1.5\cdot10^{15}\footnotemark[2]$
&$1.1\footnotemark[3]$
&$1.7\cdot10^{20}$
&$6\cdot10^{-1}\footnotemark[4]$
\\
Benzene
&$3.5\cdot10^{15}\footnotemark[2]$
&$8.8\footnotemark[3]$
&$1.7\cdot10^{20}$
&$1.5\cdot10^{-1}\footnotemark[4]$
\\
\end{tabular}
\end{ruledtabular}
\footnotetext[1]{calculated from mass loss measurement and MC simulation of precursor distribution. }
\footnotetext[2]{approximated from chamber background pressure. }
\footnotetext[3]{estimated from dimensions of undissociated precursor molecule. }
\footnotetext[4]{estimated from molar mass and density of undissociated molecule.}
\footnotetext[5]{estimated from molar mass and deposit density using the approach described in \cite{utkedensity}. (The estimated decomposed volume of Co$_2$C$_x$O$_y$ compares with 6.3$\times$10$^{-2}$ nm$^{3}$ used for the FEBID product of W(CO)$_6$ in \cite{Hoyle}). }

\end{table}

\begin{table}
\caption{
Rate fits for cross-section and residence time using the growth rate model. The fits are based on equ. \ref{eq:eq3} using the values given in table \ref{tb:fitconstants}.
}
\label{tb:fitparameters}
\begin{ruledtabular}
\begin{tabular}{@{}lccc}
precursor
&$\sigma$
&$\tau$
&$k_r\cdot t_r$
\\

&nm$^{2}$
&$\mu$s
&
\\
Co$_2$(CO)$_8$	
&$(.00495\pm.00078)\footnotemark[1]$
&$720\pm150$
&$19.7$
\\
PPE
&$(2.3\pm3.0)$
&$90\pm9$
&$111$
\\
Benzene
&$(2.09\pm2.26)\footnotemark[2]$
&$160\pm20$
&$62.5$
\\
\end{tabular}
\end{ruledtabular}
\footnotetext[1]{As a comparison, a measured $\sigma$ Benzene of 0.35nm$^2$ is reported in \cite{Kunze}.}
\footnotetext[2]{A fitted $\sigma$ Co$_2$(CO)$_8$ of 5$\times$10$^{-3}$nm$^2$ was presented in (Utke,Purruker,MNE07).}

\end{table}

In figure \ref{fig:heights2}, it can be seen that the growth rates for (600nm)$^2$ squares deposited from Co$_2$(CO)$_8$ are discontinuous, presenting similarities with hydrocarbon dissociation for low dwell times (high dependence on dwell time), whereas the growth rates at higher dwell times approach the values in figure \ref{fig:R3umCo}. We propose that this behaviour is due to co-deposition of both Cobalt carbonyl and background pressure hydrocarbons, which reach the dissociation spot by surface migration. For the 3$\mu$m$^2$ deposits, surface diffusion can be neglected, as the growing deposit represents a barrier for surface diffusion of hydrocarbon species. 

In the presence of two precursors, competitive adsorption can be modelled using:
\begin{eqnarray}
\frac{\partial n_i}{\partial t}=s_iJ_i\left(1-\frac{n_i}{n_{0,i}}-\frac{n_j}{n_{0,j}}\right)-\frac{n_i}{\tau_i}-\sigma_i fn_i
\label{eq:eq5}
\end{eqnarray}
Solving this differential equation and assuming full replenishment ($n_{out} = lim_{t\rightarrow\infty}\left[n_r(t)\right]$) yields:

\begin{subequations}
\label{eq:eq6}
\begin{eqnarray}
n_i\left(t\right)=\frac{\left(1-\frac{1}{n_{0,j}}n_j^*(t)\right)n_i^*(t)}{1-\frac{1}{n_{0,i}n_{0,j}}n_i^*(t)n_j^*(t)} \\
n_i^*\left(t\right)=n_{in,i}+\left(n_{out,i}-n_{in,i}\right)exp\left[-k_{d,i}t\right]
\end{eqnarray}
\end{subequations}
where $n_i^*$ represents the precursor surface coverage from the one-molecule model described by the constants in equation \ref{eq:eq4}. 
Comparing with equation \ref{eq:eq4}, the presence of the second precursor doesn't affect the dissociation and replenishment time constants in the model. However, the surface site occupancies are limited by the presence of the competing precursor species. The growth rate is then expressed as the sum of the growth rate contribution for each species:

\begin{eqnarray}
R_{tot}=V_1\sigma_1f t_d^{-1}\int\limits_{0}^{t_d}n_1(t)\,dt+V_2\sigma_2f t_d^{-1}\int\limits_{0}^{t_d}n_2(t)\,dt
\label{eq:eq7}
\end{eqnarray}

\begin{figure}
\begin{center}
\includegraphics[width=.45\textwidth]{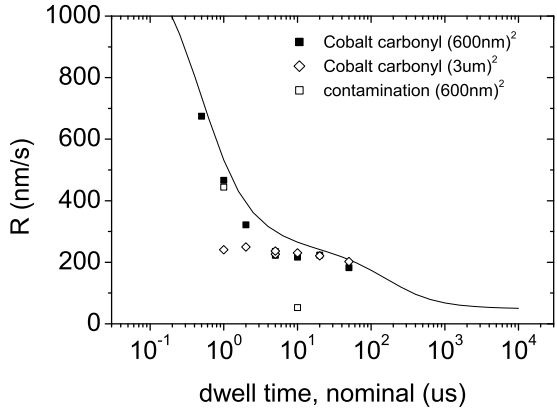}
\end{center}
\caption{Deposition rates for (600nm)$^2$ squares. The growth rates for (3$\mu$m)$^2$ structures and for contamination structures are given for comparison. The line is the model prediction as described by equation \ref{eq:eq7}. }
\label{fig:R600nmCo}
\end{figure}

\begin{figure}
\begin{center}
\includegraphics[width=.45\textwidth]{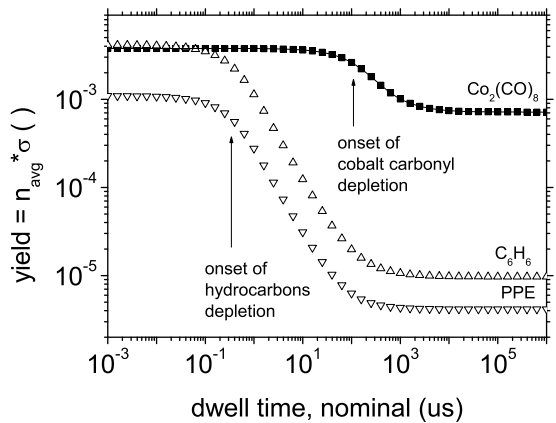}
\end{center}
\caption{Calculated yields for Cobalt carbonyl, benzene and PPE as example hydrocarbons.}
\label{fig:yields}
\end{figure}

The integrals were solved numerically and the result is shown in figure \ref{fig:R600nmCo}. The model predicts reasonably the change in growth rate observed for the (600nm)$^2$ series. Figure \ref{fig:yields} shows the variation of dissociation yield ($Y_i= \sigma_i t_d^{-1}\int\limits_0^{t_d}n_i(t)\,dt$) per electron for the Cobalt carbonyl and hydrocarbon species. As the dwell time increases, the adsorbed hydrocarbons are locally depleted, as they are dissociated efficiently, and this increases the relative contribution of Cobalt carbonyl to the deposit growth. This is similar to the switching between electron beam induced etching and deposition described in \cite{toth}, where selective depletion was achieved by varying the beam current used. 

\begin{figure}
\begin{center}
\includegraphics[width=.45\textwidth]{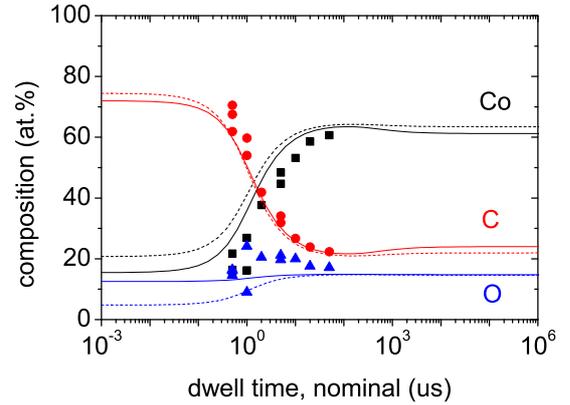}
\end{center}
\caption{Composition of 600nm squares and model prediction. Full line: PPE, dotted line: benzene.}
\label{fig:compositionfit}
\end{figure}

Under electron irradiation, the carbonyl groups of the Co$_2$(CO)$_8$ are partially dissociated from the metallic atoms and escape as volatile molecules, or are themselves dissociated under electron irradiation and get fixed in the growing deposit \cite{utkereview}. Thus, prediction of the deposit's composition is difficult, as the dissociation pathways are manifold, resulting in a complex process. However, the dissociation product of Cobalt carbonyl in our experimental conditions can be estimated using the composition of the (3$\mu$m)$^2$ deposits shown in figure \ref{fig:compositionsquar}. Based on these EDX measurements, we approximate the dissociation product of Cobalt carbonyl in our experimental conditions to Co$_2$C$_{0.61}$O$_{0.46}$. Furthermore, we assume that carbon contamination is dissociatively deposited without volatile carbon-containing fragments escaping. It is then possible to model the composition of the deposit where co-deposition of hydrocarbon species occurs, using:

\begin{eqnarray}
[Co]=\frac{Y_{Co_2(CO)_8}\cdot2}{Y_{Co_2(CO)_8}(2+0.61+0.46) + Y_{C_xO_yH_z}(x+y)}
\label{eq:eq8}
\end{eqnarray}
where Y denotes the dissociation yield, the lower term represents the number of deposited atoms per electron, and the upper term the number of deposited Co atoms per electron.	
The result is shown in figure \ref{fig:compositionfit} and compared with the composition found for the (600nm)$^2$ deposits. The two-precursor model describes the variations in composition quite well. As depletion of hydrocarbons is more pronounced, longer dwell times lead to cobalt-rich deposits. For even longer dwell times, a slight decrease of Cobalt content is predicted, as depletion of Cobalt carbonyl occurs.

In \cite{lau, utkecross, utkethermal, fernandez}, Co concentrations between 35at.\% and 90at.\% are reported. The differences are usually attributed to thermal effects by local electron-beam induced heating. However, for low beam currents ($\le$ 1nA) and SiO$_2$ as a substrate, MC simulation of electron-beam heating effects predict negligible local heating. On the other hand, the proposed co-deposition model explains a large scatter in metallic composition even at background pressures of 10$^{-6}$ mbar, lower than the one investigated here, depending on the ratio of adsorbed hydrocarbon molecules participating in the deposition.

\section{Conclusion}
The importance of contamination incorporation was shown for tungsten from WF$_6$ in \cite{hiroshima}, where plasma-cleaning of the surface and removal of surface-adsorbed hydrocarbons would lead to deposits of higher purity, as well as in \cite{lukasczyk} for the organic precursor Fe(CO)$_5$, where an ultra-high-vacuum environment increased the deposit purity. The question is discussed in the topical review in \cite{botman}. Here, it was shown how to take advantage of the concomitant deposition of residual hydrocarbons in order to tune the Cobalt content of planar deposits from Cobalt carbonyl within a range of 20 to 70 at.\%. The findings are especially relevant for using such deposits as magnetic sensing material, as the Hall resistivity depends strongly on the metallic content of Cobalt deposits, with a strong peak at around 65at.\% \cite{gabureacnanohall}.
\begin{acknowledgments}
This research is part of the EC project BioNano-Switch (043288).
\end{acknowledgments}
\section*{References}
\bibliographystyle{unsrt}
\bibliography{article}

\begin{thebibliography}{10}

\bibitem{utkereview}
I.~Utke.
\newblock Gas-assisted focused electron beam and ion beam processing and
  fabrication.
\newblock {\em Journal of vacuum science \& technology}, 26(4):1197, 2008.

\bibitem{vandorpreview}
W.~van Dorp.
\newblock A critical literature review of focused electron beam induced
  deposition journal of applied physics.
\newblock {\em Journal of Applied Physics}, 104(8):081301, 2008.

\bibitem{randolphreview}
S.~J. Randolph, J.~D. Fowlkes, and P.~D. Rack.
\newblock Focused, nanoscale electron-beam-induced deposition and etching.
\newblock {\em Critical Reviews in Solid State and Materials Sciences},
  31(3):55--89, 2006.

\bibitem{hochleitner}
G.~Hochleitner, H.~D. Wanzenboeck, and E.~Bertagnolli.
\newblock Electron beam induced deposition of iron nanostructures.
\newblock In {\em 51st International Conference on Electron, Ion, and Photon
  Beam Technology and Nanofabrication}, pages 939--944, Denver, CO, 2007. A V S
  Amer Inst Physics.

\bibitem{shimojo}
M.~Shimojo, M.~Takeguchi, K.~Mitsuishi, M.~Tanaka, and K.~Furuya.
\newblock Mechanisms of crystalline iron oxide formation in electron
  beam-induced deposition.
\newblock In {\em 19th International Microprocesses and Nanotechnology
  Conference}, pages 6247--6249, Kanagawa, JAPAN, 2006. Inst Pure Applied
  Physics.

\bibitem{lukasczyk}
T.~Lukasczyk, M.~Schirmer, H.~P. Steinruck, and H.~Marbach.
\newblock Electron-beam-induced deposition in ultrahigh vacuum: Lithographic
  fabrication of clean iron nanostructures.
\newblock {\em Small}, 4(6):841--846, 2008.

\bibitem{perentes}
A.~Perentes, G.~Sinicco, G.~Boero, B.~Dwir, and P.~Hoffmann.
\newblock Focused electron beam induced deposition of nickel.
\newblock In {\em 51st International Conference on Electron, Ion, and Photon
  Beam Technology and Nanofabrication}, pages 2228--2232, Denver, CO, 2007. A V
  S Amer Inst Physics.

\bibitem{lau}
Y.~M. Lau, P.~C. Chee, J.~T.~L. Thong, and V.~Ng.
\newblock Properties and applications of cobalt-based material produced by
  electron-beam-induced deposition.
\newblock {\em Journal of Vacuum Science \& Technology a-Vacuum Surfaces and
  Films}, 20(4):1295--1302, 2002.

\bibitem{utkespt}
I.~Utke, F.~Cicoira, G.~Jaenchen, P.~Hoffmann, L.~Scandella, B.~Dwir, E.~Kapon,
  D.~Laub, P.~Buffat, N.~Xanthopoulos, and H.~J. Mathieu.
\newblock Focused electron beam induced deposition of high resolution magnetic
  scanning probe tips.
\newblock In P.~Bernier, P.~Ajayan, Y.~Iwasa, and P.~Nikolaev, editors, {\em
  Symposium on Making Functional Materials with Nanotubes held at the 2001 MRS
  Fall Meeting}, pages 307--312, Boston, Ma, 2001.

\bibitem{utkecross}
I.~Utke, J.~Michler, Ph. Gasser, C.~Santschi, D.~Laub, M.~Cantoni, P.A. Buffat,
  C.~Jiao, and P.~Hoffmann.
\newblock Cross section investigations of compositions and sub-structures of
  tips obtained by focused electron beam induced deposition.
\newblock {\em Advanced Engineering Materials}, 7(5):323--331, 2005.

\bibitem{utkedensity}
I.~Utke.
\newblock Density determination of focused-electron-beam-induced deposits with
  simple cantilever-based method applied physics letters.
\newblock {\em Applied Physics Letters}, 88(3):031906, 2006.

\bibitem{fernandez}
A.~Fernandez-Pacheco, J.~M. De~Teresa, R.~Cordoba, and M.~R. Ibarra.
\newblock Magnetotransport properties of high-quality cobalt nanowires grown by
  focused-electron-beam-induced deposition.
\newblock {\em Journal of Physics D-Applied Physics}, 42(5):6, 2009.

\bibitem{utkethermal}
I.~Utke, T.~Bret, D.~Laub, Ph~Buffat, L.~Scandella, and P.~Hoffmann.
\newblock Thermal effects during focused electron beam induced deposition of
  nanocomposite magnetic-cobalt-containing tips.
\newblock {\em Microelectronic Engineering}, 73-74:553--558, 2004.

\bibitem{boero}
G.~Boero, I.~Utke, T.~Bret, N.~Quack, M.~Todorova, S.~Mouaziz, P.~Kejik,
  J.~Brugger, R.~S. Popovic, and P.~Hoffmann.
\newblock Submicrometer hall devices fabricated by focused
  electron-beam-induced deposition.
\newblock {\em Applied Physics Letters}, 86(4):3, 2005.

\bibitem{gabureacMNE}
M.~Gabureac, L.~Bernau, I.~Utke, O.~Kazakova, and G.~Boero.
\newblock Fabrication of sub-100nm sized hall sensors by co-febid as a tool to
  monitor dna-elongation.
\newblock {\em EuroSensors 2009, Lausanne, Switzerland}, 2009.

\bibitem{friedligis}
V.~Friedli and I.~Utke.
\newblock Optimized molecule supply from nozzle-based gas injection systems for
  focused electron- and ion-beam induced deposition and etching: simulation and
  experiment.
\newblock {\em Journal of Physics D-Applied Physics}, 42(12):11, 2009.

\bibitem{babin}
S.~Babin, M.~Gaevski, D.~Joy, M.~Machin, and A.~Martynov.
\newblock Technique to automatically measure electron-beam diameter and
  astigmatism: Beametr.
\newblock In {\em 50th International Conference on Electron, Ion, and Photon
  Beam Technology and Nanofabrication}, pages 2956--2959, Baltimore, MD, 2006.
  A V S Amer Inst Physics.

\bibitem{Kunze}
D.~Kunze, O.~Peters, and Sauerbre.G.
\newblock Polymerisation adsorbierter kohlenwasserstoffe bei
  elektronenbeschuss.
\newblock {\em Zeitschrift Fur Angewandte Physik}, 22(2):69, 1967.

\bibitem{Hoyle}
P.~C. Hoyle.
\newblock Electrical resistance of electron beam induced deposits from tungsten
  hexacarbonyl.
\newblock {\em Applied Physics Letters}, 62(23):3043, 1993.

\bibitem{toth}
M.~Toth, C.~J. Lobo, G.~Hartigan, and W.~R. Knowles.
\newblock Electron flux controlled switching between electron beam induced
  etching and deposition.
\newblock {\em Journal of Applied Physics}, 101(5), 2007.

\bibitem{hiroshima}
H.~Hiroshima, N.~Suzuki, N.~Ogawa, and M.~Komuro.
\newblock Conditions for fabrication of highly conductive wires by
  electron-beam-induced deposition.
\newblock {\em JAPANESE JOURNAL OF APPLIED PHYSICS}, 38(12B):7135--7139, 1999.

\bibitem{botman}
A.~Botman, J.~J.~L. Mulders, and C.~W. Hagen.
\newblock Creating pure nanostructures from electron-beam-induced deposition
  using purification techniques: a technology perspective.
\newblock {\em Nanotechnology}, 20(37):17, 2009.

\bibitem{gabureacnanohall}
Mihai Gabureac, Laurent Bernau, Ivo Utke, and Giovanni Boero.
\newblock Nano-hall sensors with granular co-c.
\newblock {\em Nanotechnology}, (accepted for publication), 2010.

\end{thebibliography}
\end {document}